\pdfoutput=1

\documentclass[12pt]{article}
\usepackage[latin1]{inputenc}
\usepackage{amsmath}
\usepackage{amsfonts}
\usepackage{amssymb}
\usepackage{amsthm}
\usepackage[round]{natbib}
\usepackage{algorithm}
\usepackage{algpseudocode} 
\usepackage{graphicx}
\usepackage{url}
\usepackage{subfig}
\usepackage{bm}
\usepackage{epstopdf}

\title{\bf Inference for SDE models via Approximate Bayesian Computation}
\author{Umberto Picchini \thanks{Umberto Picchini is Assistant Professor, Centre for Mathematical Sciences, Lund University, SE-22100 Lund, Sweden. email: \texttt{umberto@maths.lth.se}.}}
\date{}

\begin{document}

\def\spacingset#1{\renewcommand{\baselinestretch}%
{#1}\small\normalsize} \spacingset{1}

\maketitle

\begin{abstract}
Models defined by stochastic differential equations (SDEs) allow for the representation of random variability in dynamical systems. The relevance of this class of models is growing in many applied research areas and is already a standard tool to model e.g. financial, neuronal and population growth dynamics. However inference for multidimensional SDE models is still very challenging, both computationally and theoretically. Approximate Bayesian computation (ABC) allow to perform Bayesian inference for models which are sufficiently complex that the likelihood function is either analytically unavailable or computationally prohibitive to evaluate. A computationally efficient ABC-MCMC algorithm is proposed, halving the running time in our simulations. Focus is on the case where the SDE describes latent dynamics in state-space models; however the methodology is not limited to the state-space framework. Simulation studies for a pharmacokinetics/pharmacodynamics model and for stochastic chemical reactions are considered and a \textsc{Matlab} package implementing our ABC-MCMC algorithm is provided.

\end{abstract}


\spacingset{1.45}
\section{Introduction}

Approximate Bayesian computation (ABC) is a ``likelihood--free'' methodology which is enjoying increasingly popularity as it provides a practical approach to perform inference for models that, due to likelihood function intractability, would otherwise be computationally too challenging to be considered, see the reviews by \cite{sisson-fan(2011)} and \cite{marin-et-al(2011)}.

In this work we consider dynamical processes whose unobserved hidden state is defined by diffusion processes, that is latent dynamics are solutions to stochastic differential equations (SDEs). SDEs allow for the modelization of dynamics subject to random fluctuations, thus providing a way to represent non-deterministic behavior as observed in many applied areas, such as neuronal modelling, financial mathematics, population dynamics, pharmacokinetics/pharmacodynamics and modelling of physiological and chemical dynamics. Inference for SDEs has produced a large body of literature in the last twenty years. In this work we are primarily concerned with the estimation of unknown parameters in SDE models given noisy measurements sampled at discrete times; in particular we use Bayesian inference. As we see in a moment there are several issues affecting the success and practical implementation of (exact) Bayesian methodology for complex multidimensional SDE models, and we consider ABC methods to tackle these difficulties.

Suppose $\bm{X}_t \equiv \bm{X}_t(\bm{\psi})$ is a given (multidimensional) continuous stochastic process representing the state of a system at time $t$, depending on a vector of unknown model parameters $\bm{\psi}$ that we wish to infer from data. Also suppose $\bm{X}_t$ is the solution to a SDE which will be explicitly  introduced in section \ref{sec:SDE-models}. Depending on the application scenario the initial state of the system $\bm{X}_{t_0}$ might be known and set equal to a constant $\bm{X}_{t_0}=\bm{x}_0$, otherwise we consider it as an unknown parameter; in the latter case $\bm{\psi}$ includes $\bm{X}_0$. In general we may consider $\bm{X}_t$ as unobservable and only known up to some error $\bm{\varepsilon}_t$ that prevents exact measurement of $\bm{X}_t$ to be obtained. Therefore in correspondence to a set of $n+1$ discrete time instants $t_0 < t_1 < \cdots <  t_n$ ($t_0\geq 0$)  there is a set of ``corrupted'' versions of $\bm{X}_0,\bm{X}_1,...,\bm{X}_n$ ($\bm{X}_i\equiv \bm{X}_{{t_i}}$) denoted with $\bm{y}_0,\bm{y}_1,...,\bm{y}_n$, where the generic $\bm{y}_i$ ($\equiv \bm{y}_{{t_i}}$) is a draw from some probability distribution underlying the following error--model 
\begin{equation}
\bm{Y}_i=\bm{f}(\bm{X}_i,\bm{\varepsilon}_i), \qquad i=0,1,...,n \label{eq:error-model}
\end{equation}
where $\bm{f}(\cdot)$ is a known real-valued vector function. In many cases it is reasonable to assume that the $\bm{\varepsilon}_i$'s are independent draws from some probability distribution and therefore consider the $\{\bm{Y}_i\}$'s as conditionally independent given the $\{\bm{X}_i\}$'s. However conditional independence is not actually needed in our methodology.

For illustration purposes we consider the additive case $\bm{f}(\bm{X}_i,\bm{\varepsilon}_i)=\bm{X}_i+\bm{\varepsilon}_i$, although such choice does not affect theoretical developments. A further non-constraining assumption is to consider  $\bm{\varepsilon}_i$ as being i.i.d. with mean zero and unknown covariance matrix, e.g. diagonal covariance $\sigma^2_{\varepsilon}\bm{I}_{\mathrm{dim}(\bm{Y}_i)}$ (we always consider the case of homoscedastic variances), where in realistic experimental scenarios the value of $\sigma_{\varepsilon}$ has to be inferred from data. These assumptions on the error term are set for simplicity of exposition, and more complicated error structures can be considered without affecting theoretical considerations (e.g. correlated, state-dependent errors). Ideally we wish to make inference about $\bm{\theta}=(\bm{\psi},\sigma_{\varepsilon})$ using Bayesian methods, i.e. given $\bm{y}=(\bm{y}_0,\bm{y}_1,...,\bm{y}_n)^T$ we want to sample from the posterior density $\pi(\bm{\theta}|\bm{y})$ (here $T$ denotes transposition). However for many complex models not only is a closed form expression for $\pi(\bm{\theta}|\bm{y})$ unavailable, but also very general Markov chain Monte Carlo (MCMC) methods such as the Metropolis-Hastings algorithm may fail for a number of reasons, including e.g. difficulties in exploring the parameter space (poor mixing in the chain), multimodality in the posterior surface, difficulties in constructing adequate proposal densities for $\bm{X}_t$. Considering that for SDE models simulated trajectories are by nature highly erratic, distance from the observed data might turn unacceptably high for an MCMC algorithm based on acceptance/rejections of proposals, even if the starting values of the parameters are located in the bulk of the posterior distribution. 
Inference for SDEs is particularly difficult when either the explicit solution of the SDE or the transition density of the underlying diffusion process are unavailable. For some \textit{ad hoc} models or for sufficiently simple applications successful inferential strategies for SDE models are available (see the reviews in \cite{hurn-jeisman-lindsay(2007)} and \cite{sorensen(2004)}), however the aforementioned difficulties sometimes prevent the application of SDE models, particularly in situations where many parameters need to be estimated over multiple dimensions. Furthermore the presence of measurement error is a factor complicating the inference considerably.

In this work we propose a feasible route to approach a general class of (multidimensional) SDE models, by exploiting recent results in likelihood-free ABC methodology, circumventing the evaluation of the intractable likelihood and targeting an approximation to the posterior $\pi(\bm{\theta}|\bm{y})$. A computationally efficient ABC-MCMC algorithm is offered, exploiting the considered ABC methods and accelerating computations by 40\%--50\% in our experiments. Simulation results for a pharmacokinetics/pharmacodynamics model and stochastic chemical reactions are considered.
Likelihood-free approaches for Bayesian inference in SDE models have already been studied (e.g. in \cite{golightly2011bayesian}), potentially resulting in exact inference (except for the necessary discretisation error to perform forward simulation from the SDE): here we consider a more common scenario, where exact inference is assumed not feasible and real data $\bm{y}$ are compared with simulated data via ``summary statistics'', see section \ref{sec:abc-mcmc}. Notice that the simulation machinery we consider is MCMC and an alternative approach using sequential Monte Carlo (SMC) is implemented in the \textsc{ABC-SysBio} package \citep{liepe2010abc}, see also \cite{toni-et-al-2009} for modelling of dynamical systems via ABC-SMC. Maximum likelihood strategies using a likelihood-free approach for state-space models are considered in \cite{ionides2006inference}, \cite{breto2009time} and references therein under the umbrella of ```plug-and-play'' methods.

\section{The inferential problem for SDE models}\label{sec:SDE-models}

We are concerned with the problem of conducting Bayesian inference for the vector-valued parameter $\bm{\theta}=(\bm{\psi},\sigma_\varepsilon)$ parametrizing an SDE model, via $\bm{\psi}$, and an error model for observations contaminated with measurement error, via $\sigma_\varepsilon$.
Specifically we assume that dynamics of a given system are defined via a $d$-dimensional time-inhomogeneous (It{\^o}) stochastic differential equation for the system state $\bm{X}_t$ at time $t\geq t_0$
\begin{equation}
d\bm{X}_t = \bm{\mu}(\bm{X}_t,t,\bm{\psi})dt + \bm{\sigma}(\bm{X}_t,t,\bm{\psi})d\bm{W}_t, 
\label{eq:SDE}
\end{equation}
with $\bm{X}_0\equiv \bm{X}_{t_0}\in \mathbb{R}^d$ a known constant $\bm{x}_0$ or an unknown random quantity with density function $\pi(\bm{x}_0)$ (in the context of Bayesian statistics we can say that $\pi(\bm{x}_0)$ is the prior density of $\bm{x}_0$), and in the latter case we consider $\bm{x}_0$ as an element of $\bm{\theta}$. Here $\bm{\mu}(\cdot)$ is a $d$-dimensional real-valued vector, $\bm{\sigma}(\cdot)$ a $d\times m$ matrix and $d\bm{W}_t\sim N_m(0,\bm{I}_mdt)$ represents independent increments of an $m$--dimensional standard Brownian motion ($N_m$ is the $m$-dimensional multivariate normal distribution and $\bm{I}_m$ is the $m \times m$ identity matrix). We assume that standard regularity conditions for the existence and uniqueness of a solution for \eqref{eq:SDE} are met \citep{oksendal(2003)}. 
In most cases we assume that $\bm{X}_t$ is not observed directly, that is we consider $\bm{X}_t$ as a latent stochastic process and another process $\bm{Y}_t$ is observed instead. For example, as mentioned in the Introduction we might consider the ``error model'' $\bm{Y}_t=\bm{f}(\bm{X}_t,\bm{\varepsilon}_t)$ and the model object of study becomes
\begin{equation}
\begin{cases}
d\bm{X}_t = \bm{\mu}(\bm{X}_t,t,\bm{\psi})dt + \bm{\sigma}(\bm{X}_t,t,\bm{\psi})d\bm{W}_t\\
\bm{Y}_t=\bm{f}(\bm{X}_t,\bm{\varepsilon}_t), \quad \bm{\varepsilon}_t\sim \pi(\bm{\varepsilon}_t|\sigma_{\varepsilon}).
\end{cases}
\label{eq:state-space}
\end{equation}
When the $\{Y_t\}$ are conditionally independent given $\{X_t\}$ then \eqref{eq:state-space} is a state-space model (or a hidden Markov model), however the inferential methods here presented do not require conditional-independence to hold.

When data are measurement-error-free, i.e. have been generated by model \eqref{eq:SDE}, inference is about $\bm{\theta}\equiv \bm{\psi}$ and is based on data $\bm{x}=(x_{0,1},...,x_{0,d},...,x_{i,1},...,x_{i,d},...,\allowbreak x_{n,d})^T$,
$x_{i,j}$ being the observed value of the $j$th coordinate of $\bm{X}_t$ at time $t_i$ ($i=0,1,...,n$; $j=1,...,d$).
Otherwise inference is about $\bm{\theta}=(\bm{\psi},\sigma_{\varepsilon})$ (which might realistically include $\bm{x}_0\in\mathbb{R}^d$) and data are available through \eqref{eq:state-space} and are denoted with $\bm{y}=(y_{0,1},...,y_{0,d},...,y_{i,1},...\allowbreak ,y_{i,d},...,y_{n,d})^T$ using an obvious notation. An extended notation for partially observed systems is considered in section \ref{sec:partially-observed}. All the concepts that follow can be indifferently applied to both cases.
In the context of using the Bayesian paradigm to estimate $\bm{\theta}$, in section \ref{sec:abc-mcmc} ABC-MCMC methodology is introduced.

\section{ABC-MCMC methods}\label{sec:abc-mcmc}

For ease of notation in this section we use lower case letters to denote random variables. One of the most powerful results in MCMC theory is the Metropolis-Hastings (MH) algorithm: MH can be used to generate a Markov chain ``targeting'' a given probability distribution under rather mild regularity conditions, see \cite{brooks-et-al(2011)}. For our convenience, we intend to target the ``augmented'' posterior distribution $\pi(\bm{x},\bm{\theta}|\bm{y})$
 by using an arbitrary kernel $q((\bm{\theta}_\#,\bm{x}_\#),(\bm{x}',\bm{\theta}'))$ specified as $q((\bm{\theta}_\#,\bm{x}_\#),(\bm{x}',\bm{\theta}'))=u(\bm{\theta}'|\bm{\theta}_\#)v(\bm{x}'|\bm{\theta}')$ to propose a move from $(\bm{\theta}_\#,\bm{x}_\#)$ to $(\bm{\theta}',\bm{x}')$. As shown e.g. in \cite{sisson-fan(2011)} it is possible to simplify the MH acceptance probability to be likelihood-free as in equation \eqref{eq:acceptance-prob}:
\begin{equation}
\alpha((\bm{\theta}_\#,\bm{x}_\#),(\bm{x}',\bm{\theta}'))=1\wedge\frac{\pi(\bm{\theta}')\pi(\bm{y}|\bm{x}',\bm{\theta}')u(\bm{\theta}_\#|\bm{\theta}')}{\pi(\bm{\theta}_\#)\pi(\bm{y}|\bm{x}_\#,\bm{\theta}_\#)u(\bm{\theta}'|\bm{\theta}_\#)}.
\label{eq:acceptance-prob}
\end{equation}
where $a\wedge b = \min(a,b)$.
Now the problem of generating numerically a sufficiently accurate proposal (trajectory in our case) $\bm{x}'$ is actually a non-issue for SDE models as several approximation schemes are available \citep{kloeden-platen(1992)} and for one-dimensional SDEs it is even possible to simulate the solution exactly (\cite{beskos-et-al(2006)}). Therefore, if an $\bm{x}'$ drawn exactly from $\pi(\bm{x}'|\bm{\theta}')$ is not available we can at least produce a draw $\bm{x}'_h$ from $\pi_h(\bm{x}'|\bm{\theta}')$ where $h$ is the stepsize used in the numerical method of choice (e.g. Euler-Maruyama, Milstein, Stochastic Runge-Kutta) to discretize the SDE and obtain a numerical solution $\bm{x}'_h$, while $\pi_h(\bm{x}'|\bm{\theta}')$ is the density function for the corresponding distribution law associated with the numerical scheme. What is important to recognize is that knowledge of $\pi_h(\cdot)$ is not necessary as this gets simplified out as in \eqref{eq:acceptance-prob}, whereas the only relevant fact is having ``somehow'' generated an $\bm{x}'$ (if feasible) or an $\bm{x}'_h$, and in the latter case \eqref{eq:acceptance-prob} would depend on $h$. 
It is now clear that a feature making likelihood--free methodology attractive is that it's usually easier to simulate realisations from models \eqref{eq:SDE}--\eqref{eq:state-space} than to evaluate their likelihood functions.
As previously mentioned high rejection rates are typical in Bayesian inference for SDE models, 
when simulated trajectories are not close enough to the observed $\bm{y}$ (even when the current value of $\bm{\theta}$ lies inside the bulk of the posterior density), 
particularly for a large sample size $n$, and in such context ABC methods turn useful, as motivated in next section (although ``bridging methods'' can alleviate these issues -- see e.g. \cite{besk:papa:robe:fear:2006} -- these are difficult to generalize to multidimensional systems).  

An approximation to the previously discussed likelihood--free methodology is given by accepting proposed parameters when the corresponding simulated trajectories result sufficiently close to $\bm{y}$ according to some choice of summary statistics, a given metric and a tolerance value. This approach is at the heart of approximate Bayesian computation (ABC). If we denote with $\bm{y}_{sim}$ simulated trajectories for \textit{observable} states and with $\bm{S}(\cdot)$ a chosen vector of statistics that we apply to both $\bm{y}_{sim}$ and real data $\bm{y}$ to obtain $\bm{S}(\bm{y}_{sim})$ and $\bm{S}(\bm{y})$, it is possible to compare observed and simulated observations using some distance $|\bm{S}(\bm{y}_{sim})-\bm{S}(\bm{y})|\leq\delta$ for some tolerance $\delta>0$, or more in general using $\rho(\bm{S}(\bm{y}_{sim}),\bm{S}(\bm{y}))\leq\delta$ for some metric $\rho(\cdot)$. When such strategy is embedded into an MCMC algorithm the resulting ABC-MCMC chain targets $\pi(\bm{\theta}|\rho(\bm{S}(\bm{y}_{sim}),\bm{S}(\bm{y}))\leq\delta)$; see \cite{sisson-fan(2011)} and \cite{marin-et-al(2011)} for a general introduction to ABC methods. A frequently used approach is to consider 
\begin{equation}
\rho_{\delta}(\bm{S}(\bm{y}_{sim}),\bm{S}(\bm{y}))= \frac{1}{\delta}K\biggl(\frac{|\bm{S}(\bm{y}_{sim})-\bm{S}(\bm{y})|}{\delta}\biggr)
\label{eq:non-parametric} 
\end{equation}
where $K(\cdot)$ is a smoothing kernel density centred at $\bm{S}(\bm{y}_{sim})=\bm{S}(\bm{y})$ and $\delta$ takes the role of bandwidth. This way $\rho_{\delta}(\bm{S}(\bm{y}_{sim}),\bm{S}(\bm{y}))$ has high values when $\bm{S}(\bm{y}_{sim})$ is close to $\bm{S}(\bm{y})$.

The reasoning above can be merged with the previous likelihood-free considerations, using $\rho_{\delta}(\bm{S}(\bm{y}_{sim}),\bm{S}(\bm{y}))$ in place of $\pi(\bm{y}|\bm{x},\bm{\theta})$ to weight closeness of simulated trajectories to data in \eqref{eq:acceptance-prob}. We consider as a basic building block for the subsequent developments an ABC-MCMC algorithm discussed in \cite{sisson-fan(2011)} (there denoted ``LF-MCMC''), which is a generalization of the algorithm proposed in \cite{marjoram-et-al(2003)}. We additionally consider the case of observations perturbed via measurement error and the fact that $\delta$ itself can be considered as an unknown quantity whose appropriate value can be determined a-posteriori by examining its own Markov chain, as suggested in \cite{bortot-et-al(2007)} (see also section \ref{sec:bandwidth}). This procedure is given in Algorithm \ref{alg:lfmcmc-aug-space}.
\begin{algorithm}
\footnotesize
\caption{An ABC-MCMC algorithm with augmented state-space}
\begin{algorithmic}
\State 1. Initialization: choose or simulate $\bm{\theta}_{start}\sim \pi(\bm{\theta})$, simulate $\bm{x}_{start}\sim \pi(\bm{x}|\bm{\theta}_{start})$ and $\bm{y}_{start}\sim \pi(\bm{y}|\bm{x}_{start},\bm{\theta}_{start})$. Fix $\delta_{start}>0$ and $r=0$. Starting values are $(\bm{\theta}_r,\delta_r)\equiv(\bm{\theta}_{start},\delta_{start})$ and $\bm{S}(\bm{y}_{sim,r})\equiv \bm{S}(\bm{y}_{start})$.\\\\

    At $(r+1)$th MCMC iteration:
\State 2.  generate $(\bm{\theta}',\delta')\sim u(\bm{\theta},\delta|\bm{\theta}_r,\delta_r)$ from its proposal distribution;
\State 3.  generate $\bm{x}'\sim \pi(\bm{x}|\bm{\theta}')$ from its distribution law conditional on the $\bm{\theta}'$ from step 2; generate $\bm{y}_{sim}\sim\pi(\bm{y}|\bm{x}',\bm{\theta}')$ and calculate $\bm{S}(\bm{y}_{sim})$;
\State 4. with probability $1\wedge\frac{\pi(\bm{\theta}')\pi(\delta')K(|\bm{S}(\bm{y}_{sim})-\bm{S}(\bm{y})|/\delta')u(\bm{\theta}_r,\delta_r|\bm{\theta}',\delta')}{\pi(\bm{\theta}_r)\pi(\delta_r)K(|\bm{S}(\bm{y}_{sim,r})-\bm{S}(\bm{y})|/\delta_r)u(\bm{\theta}',\delta'|\bm{\theta}_r,\delta_r)}$ set $(\bm{\theta}_{r+1},\delta_{r+1},\bm{S}(\bm{y}_{sim,r+1})):=(\bm{\theta}',\delta',\bm{S}(\bm{y}_{sim}))$ otherwise set $(\bm{\theta}_{r+1},\delta_{r+1},\bm{S}(\bm{y}_{sim,r+1})):=(\bm{\theta}_r,\delta_r,\bm{S}(\bm{y}_{sim,r}))$; 
\State 5. increment $r$ to $r+1$ and go to step 2.
\normalsize
\end{algorithmic}
\label{alg:lfmcmc-aug-space}
\end{algorithm}
We recall that a sufficient condition for sampling \textit{exactly} from the marginal posterior $\pi(\bm{\theta}|\bm{y})$ is that the following hold simultaneously: (i) $\delta = 0$; (ii) $\bm{S}(\cdot)$ is a sufficient statistic for $\bm{\theta}$. We turn the attention at how to compensate for unattainability of conditions (i)--(ii) in section \ref{sec:summary-statistics}. 

Regarding our original inferential problem of estimating parameters for an SDE model, we consider $\bm{x}=(\bm{x}_0,\bm{x}_1,...,\bm{x}_n)^T$ as the simulated values obtained from the (analytic or numeric) solution of $\eqref{eq:SDE}$ at times $\{t_0,t_1,...,t_n\}$. When the SDE's analytic solution is not available the values in $\bm{x}$ can be obtained via discretization methods such as Euler-Maruyama, Milstein, or based on higher order It{\^o}--Taylor approximations (see \cite{kloeden-platen(1992)} for a comprehensive review). 
Should an approximated solution to \eqref{eq:SDE} be obtained, when $d=1$ we construct the (linearly) interpolated values at observational times $t_0,t_1,...,t_n$ using the finer time-grid on which the numerical discretisation is performed, and define $\bm{x}'=(\bm{x}_0,\bm{x}_1,...,\bm{x}_n)^T$ (or $\bm{x}_h'\doteq (\bm{x}_0^{(h)},...,\bm{x}_n^{(h)})^T$ for the approximated solution obtained with a stepsize $h$). For a generic value of $d$ we concatenate the interpolated values as $\bm{x}'=(x_{0,1},...,x_{0,d},...,\allowbreak x_{i,1},...,x_{i,d},...,x_{n,d})^T$, $x_{i,j}$ being the simulated value of the $j$th coordinate of $\bm{X}_t$ interpolated at time $t_i$ ($i=0,1,...,n$; $j=1,...,d$).
In next section we build on algorithm \ref{alg:lfmcmc-aug-space} and propose a new ABC-MCMC algorithm, often resulting in a dramatic reduction in computational times whenever a specific choice for the kernel $K(\cdot)$ is made.

\section{ABC-MCMC acceleration via Early--Rejection}\label{sec:early-rejection}

It turns out that for some specific choice of the kernel $K(\cdot)$ the previously described ABC-MCMC algorithm can be significantly accelerated. We simply avoid simulating from the SDE model as soon as it is known that the proposed parameter will be rejected. Under some conditions this is trivial to verify and the speed-up is achieved by simply switching the order of the calculations in the Metropolis--Hastings algorithm (but see some note of caution below): we first generate the uniform random number $\omega\sim U(0,1)$ which has to be simulated in order to evaluate whether to accept/reject the proposed parameter, then the ratio of priors and proposal densities is evaluated and when $\omega$ is larger than this ratio we can immediately reject the proposal \textit{without} simulating from the SDE model. We now proceed to specify a suitable choice for $K(\cdot)$ and then motivate how the acceleration speedup is achieved.
We always consider $K(\cdot)$ to be the uniform kernel $K(\bm{z})$ returning 1 when $\bm{z}^T\bm{Az}<c$ and 0 otherwise. In our case $\bm{z}=|\bm{S}(\bm{y}_{sim})-\bm{S}(\bm{y})|/\delta$ and $\bm{A}$ is chosen to be a $p \times p$ diagonal matrix defining the relative weighting of the parameters in the loss function (see \cite{fearnhead-prangle(2011)}). The uniform kernel is defined on a region $\bm{z}^T\bm{Az}$ bounded by a volume $c$ which is taken to be $c=V_p |\bm{A}|^{1/p}$, where $V_p=\pi^{-1}[\Gamma(p/2)p/2]^{2/p}$ (here $\pi$ denotes the mathematical constant $\pi=3.14...$). Such $c$ is the unique value producing a valid probability density function $K(\cdot)$ i.e. such that the volume of the region $\bm{z}^T\bm{Az}<c$ equals 1, see \cite{prangle(2011)}. 

Interestingly, it is possible to achieve a considerable computational acceleration (about 40--50\% in our applications) by appropriately exploiting the binary nature of such $K(\cdot)$.
The idea works as follow: by considering a uniform kernel as previously specified, the $K(\cdot)$ in the denominator of the acceptance probability always equal 1, and in practice this is necessary for the algorithm to start. Therefore the acceptance ratio simplifies to: 
\[\mathrm{ratio}=\frac{\pi(\bm{\theta}')\pi(\delta')u(\bm{\theta}_r,\delta_r|\bm{\theta}',\delta')}{\pi(\bm{\theta}_r)\pi(\delta_r)u(\bm{\theta}',\delta'|\bm{\theta}_r,\delta_r)}K(|\bm{S}(\bm{y}_{sim})-\bm{S}(\bm{y})|/\delta').\]
What is mostly relevant is that, because of the binary nature of $K(\cdot)$,
we first check whether $\omega >(\pi(\bm{\theta}')\pi(\delta')u(\bm{\theta}_r,\delta_r|\bm{\theta}',\delta'))/(\pi(\bm{\theta}_r)\pi(\delta_r)u(\bm{\theta}',\delta'|\bm{\theta}_r,\delta_r))$: if such condition is verified then we can \textit{immediately reject} $(\bm{\theta}',\delta')$ (i.e. without having to simulate from the SDE model!) as in such case $\omega$ is always larger than the ratio, regardless of the value of $K(\cdot)$. In case the check above fails, then we are required to compute $K(|\bm{S}(\bm{y}_{sim})-\bm{S}(\bm{y})|/\delta')$ and accept the proposal if $\omega\leq\mathrm{ratio}$ and reject otherwise. The procedure is coded in Algorithm \ref{alg:lfmcmc-earlyrej}.
\begin{algorithm}
\scriptsize
\caption{Early--Rejection ABC-MCMC}
\begin{algorithmic}
\State 1. Initialization: choose or simulate $\bm{\theta}_{start}\sim \pi(\bm{\theta})$, simulate $\bm{x}_{start}\sim \pi(\bm{x}|\bm{\theta}_{start})$ and $\bm{y}_{start}\sim \pi(\bm{y}|\bm{x}_{start},\bm{\theta}_{start})$. Fix $\delta_{start}>0$ and $r=0$. Starting values are $(\bm{\theta}_r,\delta_r)\equiv(\bm{\theta}_{start},\delta_{start})$ and $\bm{S}(\bm{y}_{sim,r})\equiv \bm{S}(\bm{y}_{start})$ such that $K(|\bm{S}(\bm{y}_{start})-\bm{S}(\bm{y})|/\delta_{start})\equiv 1$.\\\\

    At $(r+1)$th MCMC iteration:
\State 2.  generate $(\bm{\theta}',\delta')\sim u(\bm{\theta},\delta|\bm{\theta}_r,\delta_r)$ from its proposal distribution;
\State 3. generate $\omega \sim U(0,1)$;
\If{\[\omega>\frac{\pi(\bm{\theta}')\pi(\delta')u(\bm{\theta}_r,\delta_r|\bm{\theta}',\delta')}{\pi(\bm{\theta}_r)\pi(\delta_r)u(\bm{\theta}',\delta'|\bm{\theta}_r,\delta_r)} \quad (=\text{``ratio''})\]}
\State $(\bm{\theta}_{r+1},\delta_{r+1},\bm{S}(\bm{y}_{sim,r+1})):=(\bm{\theta}_r,\delta_r,\bm{S}(\bm{y}_{sim,r}))$; \Comment{(proposal early-rejected)}
\Else{ generate $\bm{x}'\sim \pi(\bm{x}|\bm{\theta}')$ conditionally on the $\bm{\theta}'$ from step 2; generate $\bm{y}_{sim}\sim\pi(\bm{y}|\bm{x}',\bm{\theta}')$ and calculate $\bm{S}(\bm{y}_{sim})$;} 
   \If{$K(|\bm{S}(\bm{y}_{sim})-\bm{S}(\bm{y})|/\delta')=0$} 
      \State $(\bm{\theta}_{r+1},\delta_{r+1},\bm{S}(\bm{y}_{sim,r+1})):=(\bm{\theta}_r,                
      \delta_r,\bm{S}(\bm{y}_{sim,r}))$ \Comment{(proposal rejected)}
    \ElsIf{$\omega\leq \mathrm{ratio}$} 
      \State $(\bm{\theta}_{r+1},\delta_{r+1},\bm{S}(\bm{y}_{sim,r+1})):=(\bm{\theta}',   
      \delta',\bm{S}(\bm{y}_{sim}))$ \Comment{(proposal accepted)}
     \Else
     \State $(\bm{\theta}_{r+1},\delta_{r+1},\bm{S}(\bm{y}_{sim,r+1})):=(\bm{\theta}_r,        
      \delta_r,\bm{S}(\bm{y}_{sim,r}))$ \Comment{(proposal rejected)}
      \EndIf

\EndIf
\State 4. increment $r$ to $r+1$ and go to step 2.
\end{algorithmic}
\label{alg:lfmcmc-earlyrej}
\normalsize
\end{algorithm}
This simple modification brings considerable benefits, most of the times at zero cost (but see below for exceptions), particularly for ABC algorithms where typically only a moderately low fraction of draws are accepted, e.g. \cite{sisson-fan(2011)} and \cite{fearnhead-prangle(2011)} report results obtained with a 1\% acceptance rate. As a result the possibility to avoid simulating from the SDE model comes with obvious computational benefits. 
The idea itself is not new, and has been considered in e.g. \cite{beskos-et-al(2006)} and more recently in \cite{solonen-et-al(2012)}: in particular the latter coined the expression ``early--rejection'', which we employ. However to the best of our knowledge early--rejection has not been previously considered within the ABC framework.
Notice that we always generate proposals for $\bm{\theta}$ and $\delta$ independently and therefore we could
also write $u(\bm{\theta},\delta|\bm{\theta}_r,\delta_r) = u_1 (\bm{\theta}|\bm{\theta}_r)u_2(\delta|\delta_r)$ with $u_1(\cdot)$ and  $u_2(\cdot)$ the corresponding proposal distributions.
Also consider that the acceptance probabilities in algorithms \ref{alg:lfmcmc-aug-space}--\ref{alg:lfmcmc-earlyrej} have simpler expressions when the proposal distribution $u(\cdot)$ is ``symmetric'', i.e. is such that $u(\bm{\theta}_r,\delta_r|\bm{\theta}',\delta')=u(\bm{\theta}',\delta'|\bm{\theta}_r,\delta_r)$. In our simulations we always use the adaptive Metropolis random walk with Gaussian increments proposed in \cite{haario-et-al(2001)}, and in such case the ratio in step 3 of algorithm \ref{alg:lfmcmc-earlyrej} becomes $(\pi(\bm{\theta}')\pi(\delta'))/(\pi(\bm{\theta}_r)\pi(\delta_r))$. However notice that Algorithm \ref{alg:lfmcmc-earlyrej} is not uniformly faster than Algorithm \ref{alg:lfmcmc-aug-space}, i.e. the ``early-rejection check'' might actually slow-down the performance in some cases (albeit by a negligible factor corresponding to the time needed to evaluate the ratio of prior densities): for example suppose $u(\cdot)$ is symmetric and suppose uniform priors are placed on all unknowns (including $\delta$), then the early-rejection check in algorithm \ref{alg:lfmcmc-earlyrej} simplifies to ``$\text{If }\omega>1 \text{ then }$ [...]'', condition
which is of course never verified as $\omega \sim U(0,1)$ and thus early-rejection never takes place.

We now have to choose the summary statistics $\bm{S}(\cdot)$. This is a problem of paramount importance as it has been emphasized (e.g. in \cite{fearnhead-prangle(2011)}) that whereas the shape of the kernel $K(\cdot)$ has negligible impact on the statistical performance of ABC methods, on the opposite side it is essential to choose $\bm{S}(\cdot)$ appropriately.

\subsection{Summary statistics}\label{sec:summary-statistics}

We consider recent developments allowing to build informative summary statistics, ideally ``nearly sufficient'', although in practice we cannot quantify how close we get to achieve sufficiency. Recently \cite{fearnhead-prangle(2011)} proposed an ABC methodology making the determination of summary statistics almost automatic, by exploiting a classic result holding for least-squares type of loss functions. That is they proved that when choosing $\bm{S}(\bm{y})=E(\bm{\theta}|\bm{y})$, the minimum loss, based on inference using the ABC posterior, is achieved as $\delta \rightarrow 0$ by $\hat{\bm{\theta}}=E(\bm{\theta}|\bm{S}(\bm{y}))$. Therefore in the limit as $\delta \rightarrow 0$ the posterior mean from the ABC output is $E(\bm{\theta}|\bm{S}(\bm{y}))=E(\bm{\theta}|\bm{y})$, i.e. the true posterior mean.
Thanks to such a strong result a linear regression approach is considered in \cite{fearnhead-prangle(2011)} to determine the summary statistics, that is for the $j$th parameter ($j=1,...,p$) the following linear regression model is built 
\begin{equation}
\theta_j=E(\theta_j|\bm{y})+\xi_i=\beta_0^{(j)}+\beta^{(j)}\eta(\bm{y})+\xi_j, \qquad j=1,...,p
\label{eq:linear-regression}
\end{equation} 
where $\beta^{(j)}$ is a vector of unknown coefficients and $\xi_j$ is a mean-zero random term with constant variance. Each of the $p$ regression models is estimated separately by least squares and the corresponding fit $\hat{\beta}_0^{(j)}+\hat{\beta}^{(j)}\eta(\bm{y})$ provides an estimate for $E(\theta_j|\bm{y})$ which can then be used as a summary statistic, because of the optimality property of $\bm{S}(\bm{y})\equiv E(\bm{\theta}|\bm{y})$. 

In our applications we consider the case where $\eta(\bm{y})$ is the time-ordered sequence of measurements, e.g. for  $\bm{y}_i\in  \mathbb{R}^d$ we set $\eta(\bm{y})=(y_{0,1},...,y_{0,d},...,y_{i,1},...,y_{i,d},...,\allowbreak y_{n,d})$, $y_{i,j}$ being the observed value of the $j$th coordinate of $\bm{X}_t$ at time $t_i$ ($i=0,1,...,n$; $j=1,...,d$). Of course alternative choices could be considered for $\eta(\bm{y})$, particularly for large $n$, e.g. change-points \citep{iacus2011estimation}.
Therefore, when e.g. $d=1$, our summary statistic function $S(\cdot)$ for $\theta_j$ when applied to $\bm{y}$ becomes $\hat{\beta}_0^{(j)}+\hat{\beta}^{(j)}_1y_0+\hat{\beta}^{(j)}_2y_1+...+\hat{\beta}^{(j)}_{n+1}y_n$. 
In addition to multivariate linear regression we have also experimented with Lasso (\cite{tibshirani-1996}).
We determine the penalty-parameter for the Lasso by cross validation, and the regression coefficients corresponding to the ``optimal'' penalty (as returned by the \texttt{glmnet} software \citep{friedman-hastie-tibshirani-2010}, available for both \textsc{R} and \textsc{Matlab}) are used for prediction purposes.

\subsection{Partially observed systems}\label{sec:partially-observed}

Application of ABC-MCMC methodology to handle partially observed systems is straightforward. With ``partially observed'' we mean an experiment for which \textit{not all} the $d$ coordinates of the multidimensional process $\{\bm{X}_t\}$ are observed at any sampling time $t_i$. The previously introduced methodology does not require modifications to handle such scenario, where in general at a given time $t_i$ the observed $\bm{y}_i$ has dimension $d_i\leq d$ where $d_i$ is the number of observed coordinates at time $t_i$, and algorithms \ref{alg:lfmcmc-aug-space}--\ref{alg:lfmcmc-earlyrej} require, same as before, the simulation of a draw $\bm{x}'$ having $d\times n$ elements. However this time a smaller vector of artificial data $\bm{y}_{sim}$ having the same dimensions as $\bm{y}$ is generated, i.e. $\dim(\bm{y}_{sim,i})=\dim(\bm{y}_i)=d_i$.

\section{Computational issues}\label{sec:computational-issues}

\subsection{Determination of a training region}\label{sec:training-region}

Success in the application of the presented methods is affected by the identification of a ``training region'' for $\bm{\theta}$. In practice this means that for the previously described methodology to work reasonably well, when knowledge about $\bm{\theta}$ is scarce, we should first implement a pre-ABC-MCMC simulation, that is a ``pilot study'' aiming at identifying a region of the parameter space enclosing non-negligible posterior mass. \cite{fearnhead-prangle(2011)} suggested a schedule for their semi-automatic ABC: i) Perform a pilot run of ABC to determine a region of non-negligible posterior mass on which to place a prior to be used for the rest of the analysis;
ii)``Training'': simulate sets of parameter values and artificial data from the prior determined in (i);
iii) use the simulated sets of parameter values and artificial data to estimate the summary statistics;
iv) re-run an ABC algorithm with the chosen summary statistics and prior densities determined in (i).
Step (i) is optional, when from previous knowledge (e.g. literature) information about $\bm{\theta}$ can be encoded into an informative prior. In the following we consider using informative priors, thus skipping (i) and consider as ``training region'' the support of the postulated priors. However the identification of a suitable training region without having to run first a full ABC simulation is an important research question, also because the identification of priors as from (i) is affected by a questionable double use of data (\cite{berger2006case}).

\subsection{Bandwidth selection}\label{sec:bandwidth}

The parameter determining the accuracy of inference via ABC is the bandwidth $\delta$, whose appropriate value must be chosen carefully as a trade-off between accuracy in approximating the target distribution (low $\delta$) and the ability for the Markov chain Monte Carlo to mix reasonably well, i.e. to explore the support of the posterior distribution. 
Of course $\delta$ can be chosen by trial and error, that is by identifying the smallest $\delta$ producing a satisfactory enough acceptance rate. 
\cite{bortot-et-al(2007)} propose an augmented state-space approach sampling from the ABC posterior of $(\bm{\theta},\delta)$. As shown in algorithm \ref{alg:lfmcmc-earlyrej}, we need to choose a starting value for $\delta$, its prior distribution and a generation mechanism for new values of $\delta$ at the $r$th iteration of the ABC-MCMC step. We favour small values of $\delta$, for statistical efficiency, while occasionally allowing for the generation of a larger $\delta$ to avoid the problem of poor mixing in the chain. With these requirements in mind we always set for $\pi(\delta)$ a (truncated) exponential distribution with mean $\lambda$ on the support $[0,\delta_{\mathrm{max}}]$, and at the $r$th iteration of ABC-MCMC a proposal for $\delta$ is generated via Metropolis random walk with Gaussian increments.  Here $\delta_{\mathrm{max}}$ is an application specific threshold we impose on $\delta$ to prevent its proposed draws to attain unnecessarily large values, which would in any case be removed from the final analysis as explained below. We have to take into account the need to produce long chains, as some of the generated draws are filtered out from the simulation output, namely we discard all draws in $\{\bm{\theta}_r;\delta_r\geq\delta^*\}$, where $\delta^*$ is chosen as described in next sections. In tempered-transitions \citep{neal1996} proposals for a (multimodal) target density $f(\cdot)$ are produced by crossing the supports of similar densities from which it is easier to simulate, e.g. $f^{1/\tau}$ where $\tau>0$ is a sequence of ``temperatures'' smoothing peaks and valleys on the surface of $f(\cdot)$. In our case using a variable $\delta$ has the same goal, namely approximately sample from the (exact but unavailable) posterior corresponding to $\delta=0$ by considering a small $\delta>0$ while occasionally using larger $\delta$'s to ease acceptance (and thus exploration of the posterior surface).

\section{Applications}

In order to ease the application of the methodology introduced in this work the \texttt{abc-sde} \textsc{Matlab} package has been created, see \url{http://sourceforge.net/projects/abc-sde/}. Although ABC finds it \textit{raison d'\^{e}tre} in the analysis of otherwise intractable models, in the next two sections we consider models that could be treated using exact inferential methods, e.g. sequential Monte Carlo \citep{gordon2001sequential} embedded within some MCMC strategy for parameter estimation, such as particle MCMC methods (PMCMC, \cite{andrieu2010particle}, see also \cite{d.wilkinson(2012)} for an accessible introduction) or a double-SMC approach as in $\mathrm{SMC}^2$ \citep{chopin2012smc2}. It is of interest to look at ABC-MCMC performance for such examples and compare it against exact inference whenever possible, therefore for the simpler example in section 6 a comparison against PMCMC is presented. A note of caution is necessary regarding the comparison of timings, as while ABC methods are influenced by the specific setup chosen
for the tolerance $\delta$, performance of sequential Monte Carlo methods (such as PMCMC) in terms of computational time depend on the number of chosen particles. Thus it is difficult to make a fair comparison between ABC and exact methods.

\subsection{Theophylline pharmacokinetics}\label{ex:theophylline}

Here we consider the Theophylline drug pharmacokinetics, which has often been studied in literature devoted to longitudinal data modelling with random parameters (mixed--effects models). We follow a setup as similar as possible to \cite{Pinheiro_Bates_1995} and \cite{Donnet_Samson_2008}, although our model is not a mixed-effects one.
We denote with $X_t$ the level of Theophylline drug concentration in blood at time $t$.
Consider the following non-authonomous SDE model:
\begin{equation}
dX_t = \biggl(\frac{Dose\cdot K_a \cdot K_e}{Cl}e^{-K_a t}-K_eX_t\biggr)dt + \sigma dW_t, 
\label{eq:theophylline-sde}
\end{equation}
where $Dose$ is the known drug oral dose received by a subject, $K_e$ is the elimination rate constant, $K_a$ the absorption rate constant, $Cl$ the clearance of the drug and $\sigma$ the intensity of intrinsic stochastic noise.
The experimental design for a single hypothetical subject consider nine blood samples taken at 15 min, 30 min, 1, 2, 3.5, 5, 7, 9 and 12 hrs after dosing. The drug oral dose is chosen to be 4 mg.
At time $t_0=0$ the drug is administered and therefore at $t_0$ the drug concentration in blood is zero, i.e. $X_0=x_0=0$; thus it is reasonable to assume the SDE initial state to be known and as such will not be estimated. The error model is assumed to be linear, $y_i=X_i+\varepsilon_i$ where the $\varepsilon_i\sim N(0,\sigma_{\varepsilon}^2)$ are i.i.d., $i=1,...,9$. Inference is based on data $\{y_1,y_2,...,y_9\}$ collected at times $t_1= 15$ min, $t_2=30$ min, etc. Parameters of interest are $(K_e,K_a,Cl,\sigma,\sigma_{\varepsilon})$ however in practice to prevent the parameters from taking unrealistic negative values their natural logarithm is considered and we set $\bm{\theta} =(\log K_e,\log K_a,\log Cl,\log \sigma,\log \sigma_{\varepsilon})$.

Equation \eqref{eq:theophylline-sde} is linear in the state variable and a solution is readily available (see \cite{kloeden-platen(1992)}), therefore data have been simulated exactly. For data generation we used $(\log K_e,\log K_a,\log Cl,\log \sigma,\log \sigma_{\varepsilon})
=(-2.52,0.40,-3.22,\log \sqrt{0.2},\log \sqrt{0.1})$ and interpolated the obtained trajectory at sampling times to get $n=9$ values for process $\{X_i\}$ which we perturbed via the error model with $\sigma_{\varepsilon}^2=0.1$ to obtain the $y_i$'s.
However in order to check the performance of our inferential algorithm in a typical scenario where closed-form solutions are unavailable, in the several steps of ABC we always approximate the SDE solution via Euler-Maruyama on a fine time--grid obtained by dividing each interval $[t_i,t_{i+1}]$ in 20 sub-intervals of equal length, as in \cite{Donnet_Samson_2008}.

\subsubsection*{ABC MCMC with multivariate linear regression and comparison with exact Bayesian inference}

\begin{figure}[h]
\centering
\subfloat[Subfigure 1 list of figures text][$\log K_e$ vs bandwidth]{
\includegraphics[scale=0.185]{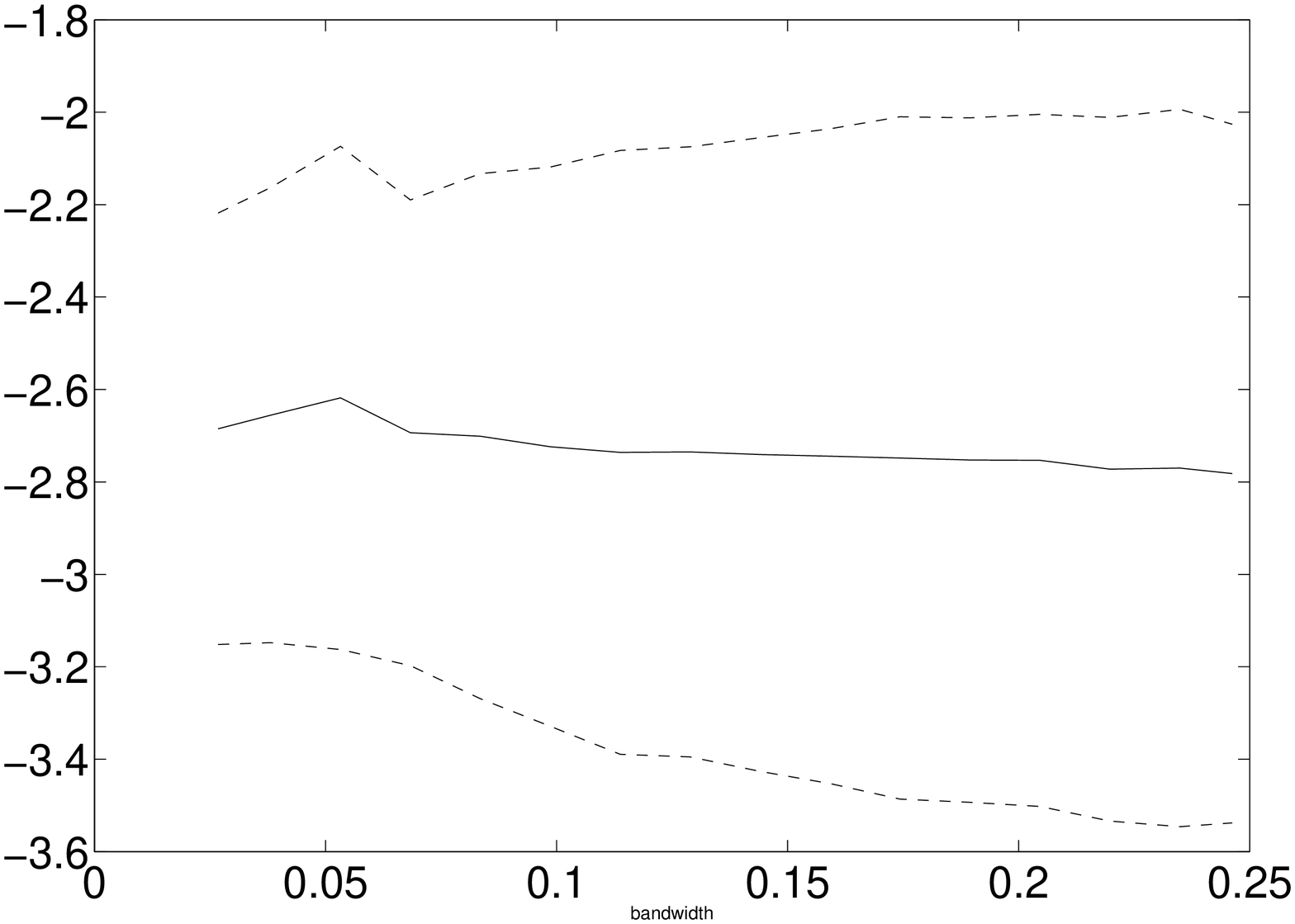}}
\quad
\subfloat[Subfigure 2 list of figures text][$\log K_a$ vs bandwidth]{
\includegraphics[scale=0.185]{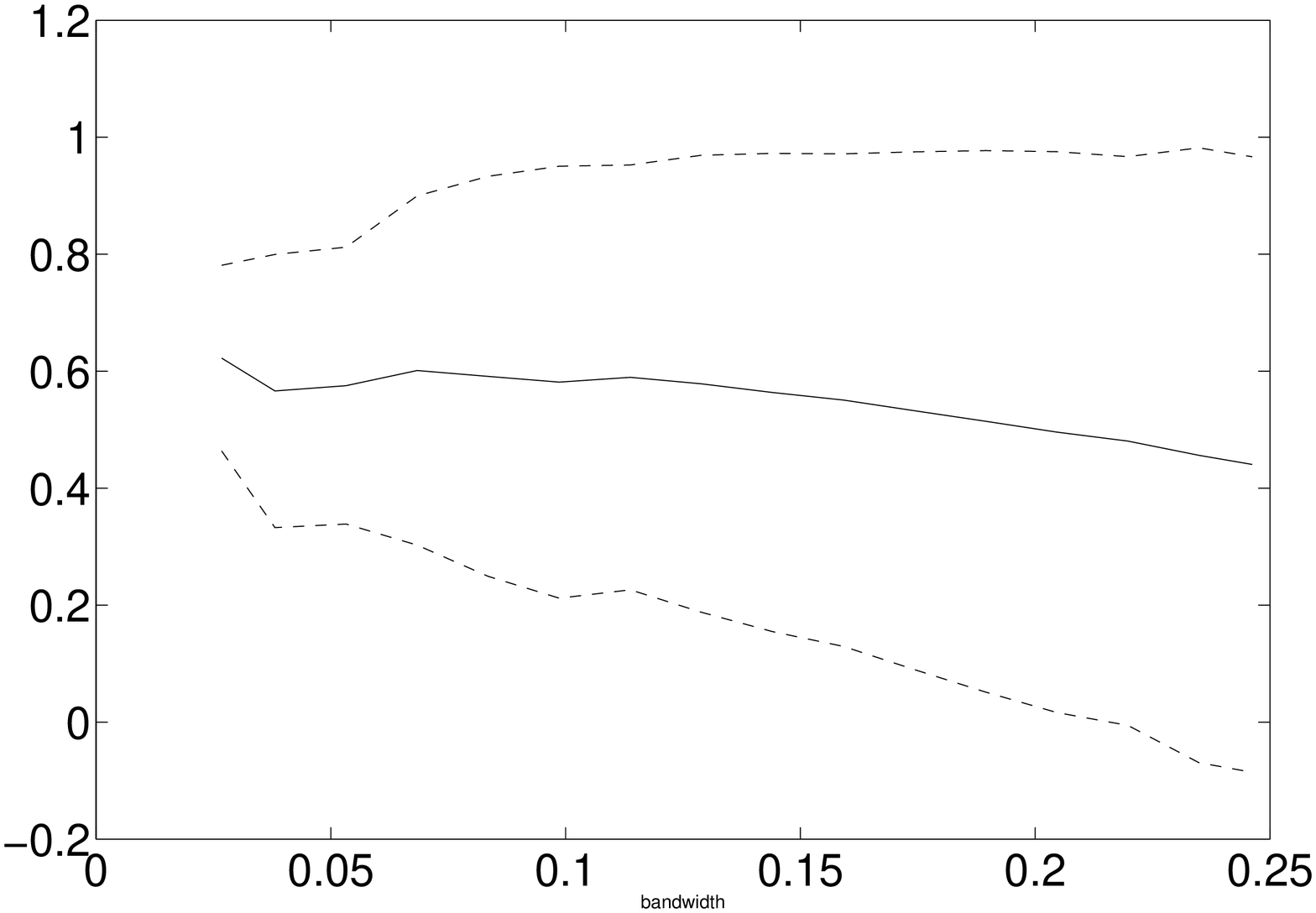}}
\quad
\subfloat[Subfigure 2 list of figures text][$\log Cl$ vs bandwidth]{
\includegraphics[scale=0.185]{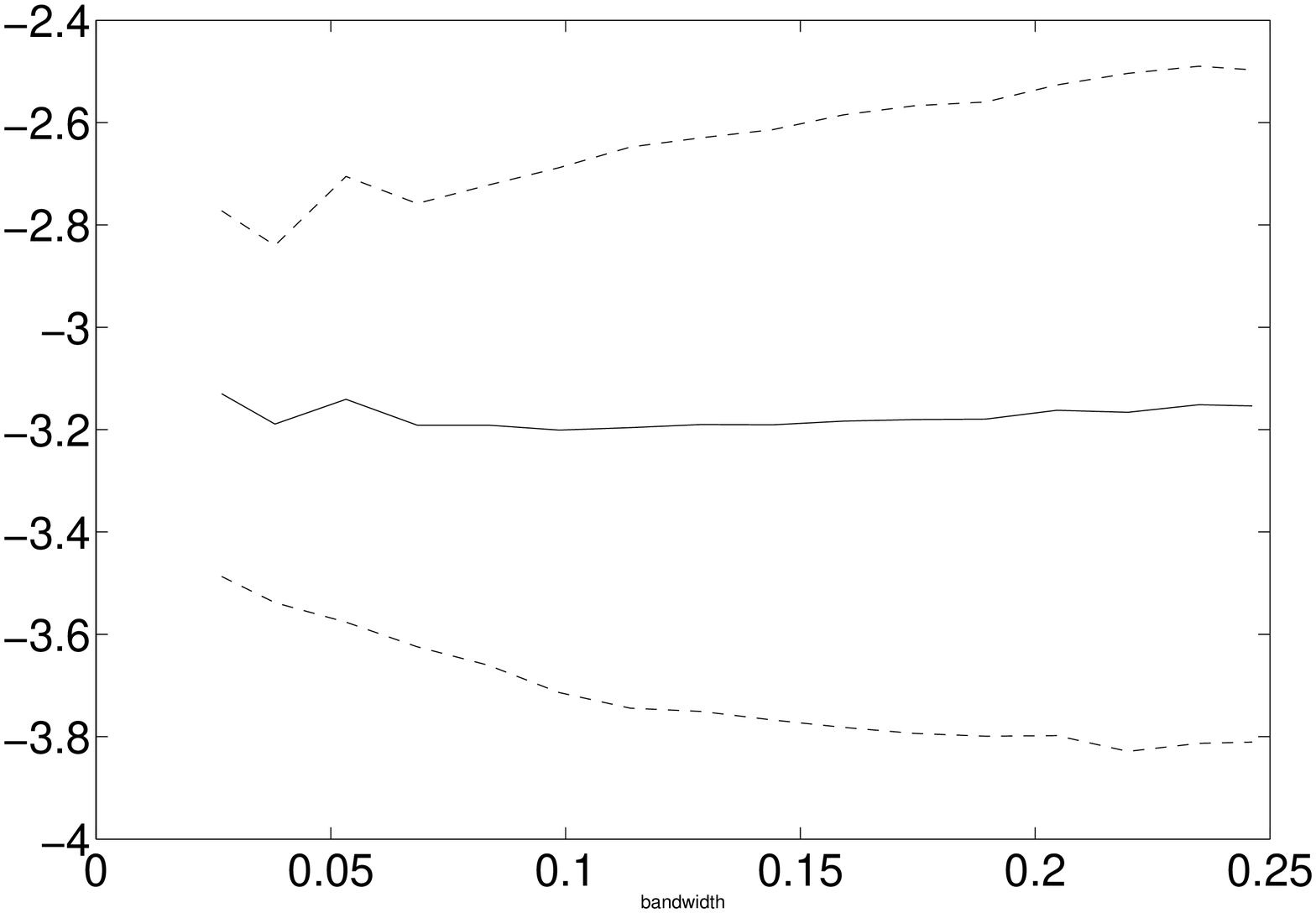}}\\
\caption{\footnotesize{Theophylline example: a subset of marginal posterior means from ABC MCMC [$\pm 2$ SD] vs bandwidth (bandwiths are on the abscissas).}}
\label{fig:parameters-vs-bandw}
\end{figure}

We consider the following priors for the unknowns of interest:
$\log K_e \sim N(-2.7,0.6^2)$, $\log K_a \sim N(0.14,0.4^2)$, $\log Cl\sim N(-3,0.8^2)$, $\log \sigma\sim N(-1.1,0.3^2)$ and $\log \sigma_{\varepsilon}\sim N(-1.25,0.2^2)$, where $N(a,b)$ denotes the Gaussian distribution with mean $a$ and variance $b$.
Prior to starting ABC-MCMC values for $\bm{S}(\cdot)$ must be obtained: we use the priors to simulate \texttt{simpar = $1000\times n=9000$} parameters and corresponding artificial observation, then use model \eqref{eq:linear-regression} to fit separately each set of simulated parameters $\theta_j$.
The prior means are used as starting value for $\bm{\theta}$ in ABC-MCMC. Algorithm \ref{alg:lfmcmc-earlyrej} is run for 3,000,000 iterations by assigning the bandwidth $\delta$ a truncated exponential prior with mean $\lambda=0.07$ and imposing a maximum allowed bandwidth of $\delta_{\mathrm{max}}=0.25$ (remember this is not $\delta^*$, which we always determine a-posteriori from the output of ABC-MCMC, see section \ref{sec:bandwidth}). By using the adaptive Metropolis random walk of \cite{haario-et-al(2001)} we achieve an 8.5\% acceptance rate, which we found a reasonable compromise between ABC accuracy and the ability to explore the surface of the approximate posterior. Notice that in ABC a high acceptance rate is not an indicator of good algorithmic performance \textit{per-se}, as it is always possible to increase the acceptance rate by enlarging $\delta$ at the expenses of statistical accuracy, for example \cite{fearnhead-prangle(2011)} obtained good results with a 1\% acceptance rate. 
Of the 3,000,000 generated draws the first 125,000 have been discarded (burn-in) then every 50th draw was retained (i.e. we used a \textit{thinning} equal to 50), this is either to diminish the strong correlation between subsequent values, which is typical in ABC when using sufficiently small $\delta$'s, and to save computer memory. In the spirit of \cite{bortot-et-al(2007)} we plot the posterior means and 95\% confidence bands for the chain of each parameter against the bandwidth $\delta$, see Figure \ref{fig:parameters-vs-bandw}. 
The purpose of these plots is to find a region $\{\delta;\delta<\delta^*\}$ where the posterior means are reasonably smooth and then filter out all draws generated from values of $\delta$ outside such region, i.e. consider for the final results only those draws $\{\bm{\theta}_r\}_{r=1,2,...}$ having $\{\bm{\theta}_r;\delta_r<\delta^*\}$. Too small values for $\delta^*$ would provide unreliable results for the posterior means as (i) these would result very variable (e.g. for $\delta<0.05$, see e.g. the panel for $\log K_e$), by being computed on a very limited number of draws or (ii) because for very small $\delta$'s the chain does not mix well (this is discussed further in section \ref{sec:stoch-kin-net}), while too large values would return very biased results. Therefore a trade-off is necessary, namely we select a small $\delta^*$ while retaining a long enough sequence of draws to allow accurate posterior inference. It is useful that we can make such choice retrospectively although this implies the need to produce long chains to balance such filtering procedure. By choosing $\delta^*=0.09$ we are left with about 5,700 draws.
Finally the posterior means from the selected draws and 95\% central posterior intervals are given in Table \ref{tab:theophylline-estimates} and results seem encouraging. We used the R \texttt{coda} package \citep{coda} in order to check for the convergence of the algorithm by computing the effective sample size (ESS) for each parameter, see Table \ref{tab:theophylline-estimates} also reporting the percentage of ESS in relation to the length of the chain, i.e. 5,700.
In terms of computational time it required about 1.8 hrs to run the MCMC part of algorithm \ref{alg:lfmcmc-earlyrej}, that is generating the 3,000,000 draws, using a \textsc{Matlab} R2011a code running on a Intel Core i7-2600 CPU 3.40 GhZ with 4 GB RAM. Using algorithm \ref{alg:lfmcmc-aug-space} it required about 3.2 hrs, therefore in this example our early--rejection approach produced a computational acceleration of about 44\%.

\begin{table}
\centering \scriptsize
\begin{tabular}{lccccc}
\hline
\hline
True parameters & $K_e = 0.080$ & $K_a = 1.492$ & $Cl=0.040$ & $\sigma=0.450$ &  $\sigma_{\varepsilon}=0.316$\\ 
\hline
Method: & {} & {} & {} & {} & {}\\
{ABC with lin. regr.} & 0.069 [0.039, 0.110]  & 1.801 [1.314, 2.479] & 0.041 [0.026, 0.062] & 0.301 [0.188, 0.471] & 0.288 [0.203, 0.392]\\ 
{\hspace{1.4cm}ESS}                    & 180.6 (3.2\%) &  194.0 (3.4\%) & 195.7 (3.4\%) & 307.5 (5.4\%) & 120.2 (2.11\%)\\ 
{ABC with lasso.} & 0.104 [0.064, 0.161]  &  1.946 [1.607, 2.475]  &  0.049 [0.036, 0.067] &   0.325 [0.289, 0.374]  &  0.222 [0.210, 0.237]\\
{\hspace{1.4cm}ESS} & 152.5 (2.7\%) & 139.5 (2.4\%) & 92.6 (1.6\%) & 113.0 (2\%) & 327.3 (5.7\%) \\  
{PMCMC, $K=100$} & 0.087 [0.051, 0.132] &   1.986 [1.514, 2.609]  &  0.048 [0.030, 0.069] &   0.332 [0.285, 0.387]  &  0.225 [0.208, 0.243]\\
{\hspace{1.4cm}ESS} & 282.6 (5.0\%) & 305.5 (5.4\%) & 287.2 (5.0\%) & 306.6 (5.4\%) & 312.3 (5.5\%)\\
{PMCMC, $K=500$} & 0.084 [0.045, 0.138] & 1.854 [1.339, 2.553] & 0.048 [0.027, 0.073] & 0.319 [0.189, 0.528] & 0.285 [0.199, 0.407]\\
{\hspace{1.4cm}ESS} & 308.5 (5.4\%) & 360.6 (6.3\%) & 297.7 (5.2\%) & 332.2 (5.8\%) & 339.8 (5.9\%)\\
\hline
\end{tabular}
\caption{\footnotesize{Theophylline example: posterior means from the ABC-MCMC output and 95\% central posterior intervals when ABC statistics are computed via linear regression and lasso. Last four lines report inference obtained via particle MCMC (PMCMC) using $K$ particles. To ease comparison between methods the ESS are all computed on samples having equal length.}}
\label{tab:theophylline-estimates} 
\end{table}

\begin{figure}[h]
\centering
\subfloat[Subfigure 1 list of figures text][$\log K_e$]{
\includegraphics[scale=0.185]{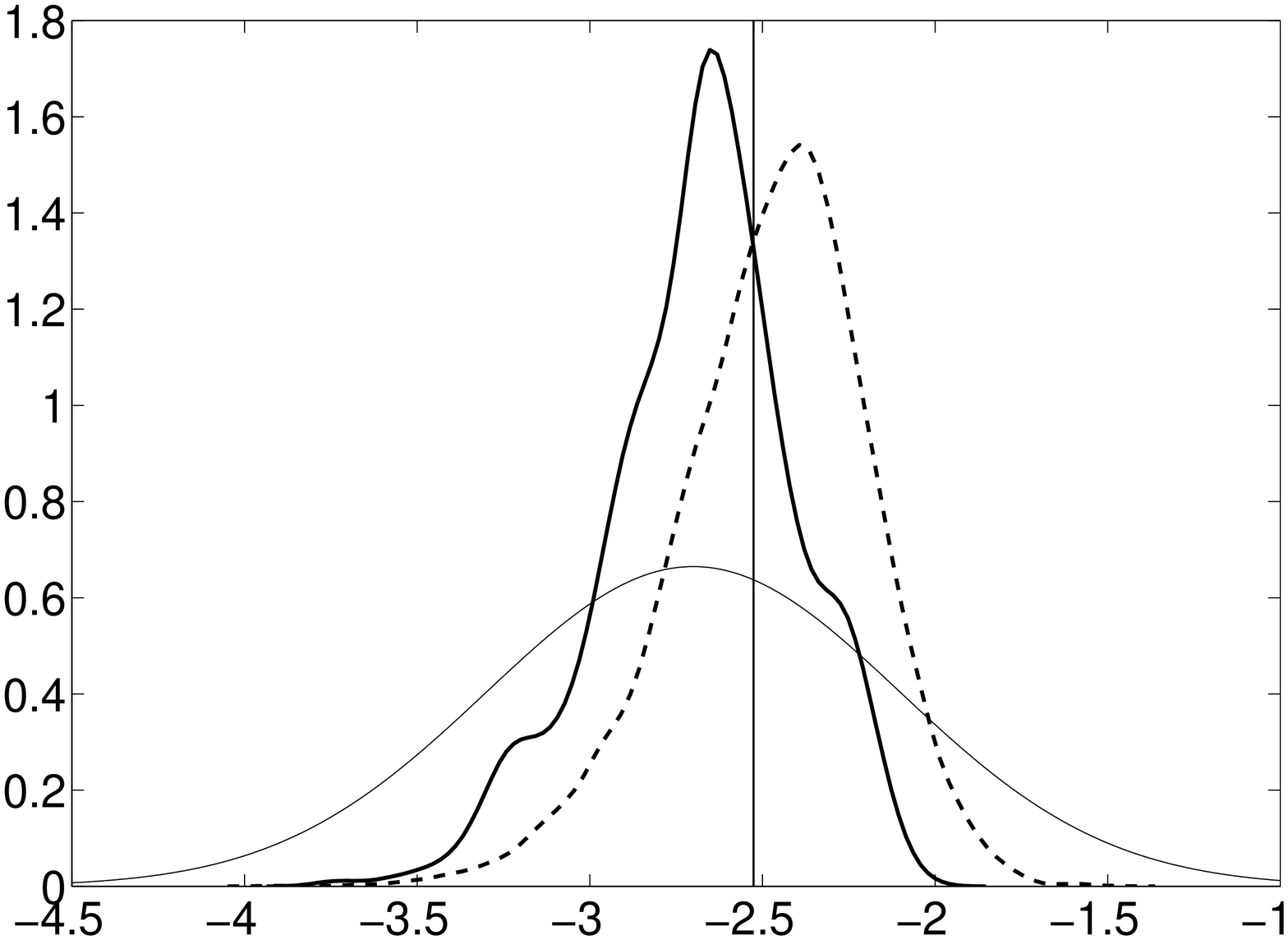}}
\quad
\subfloat[Subfigure 2 list of figures text][$\log K_a$]{
\includegraphics[scale=0.185]{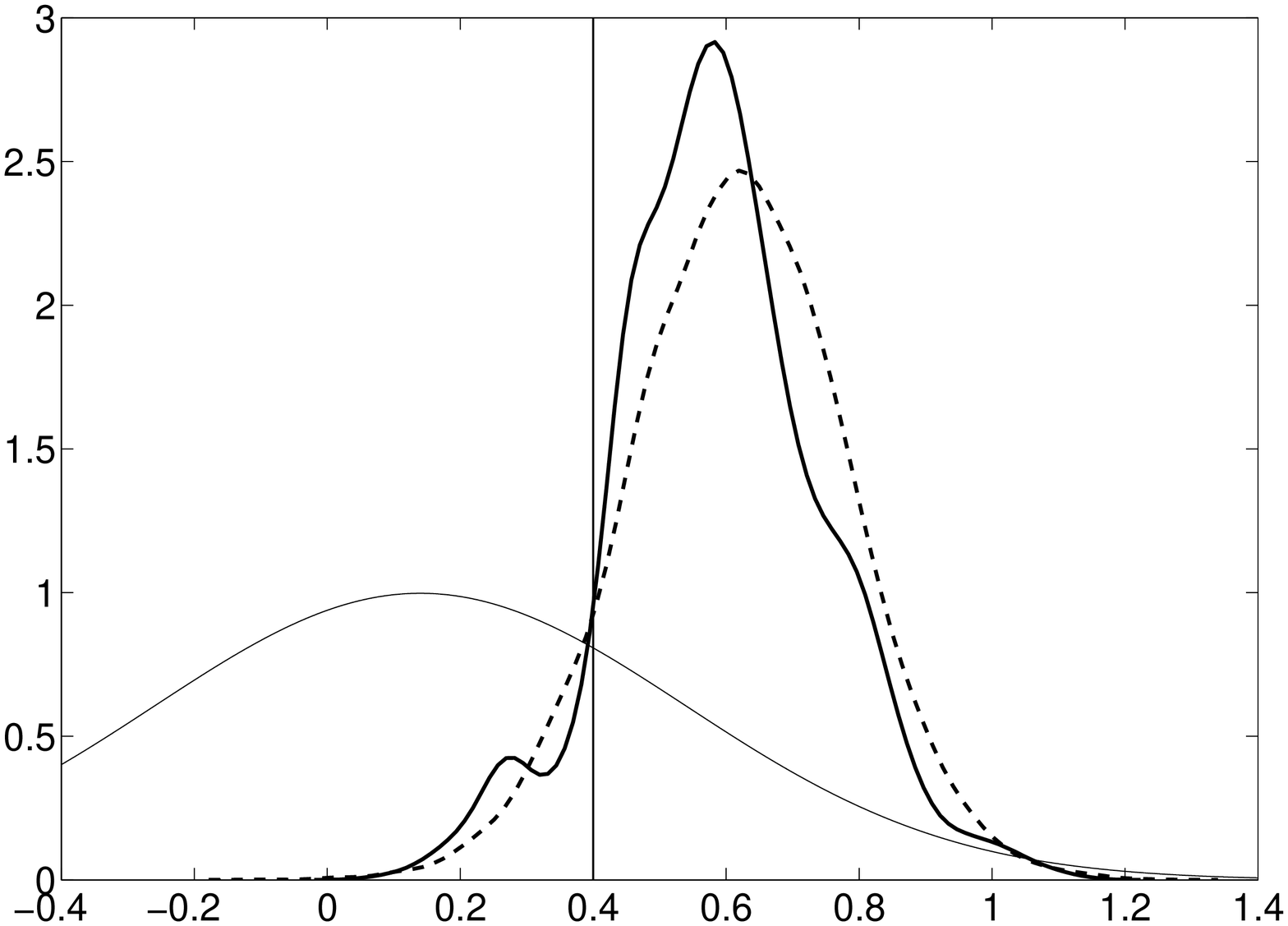}}
\quad
\subfloat[Subfigure 1 list of figures text][$\log Cl$]{
\includegraphics[scale=0.185]{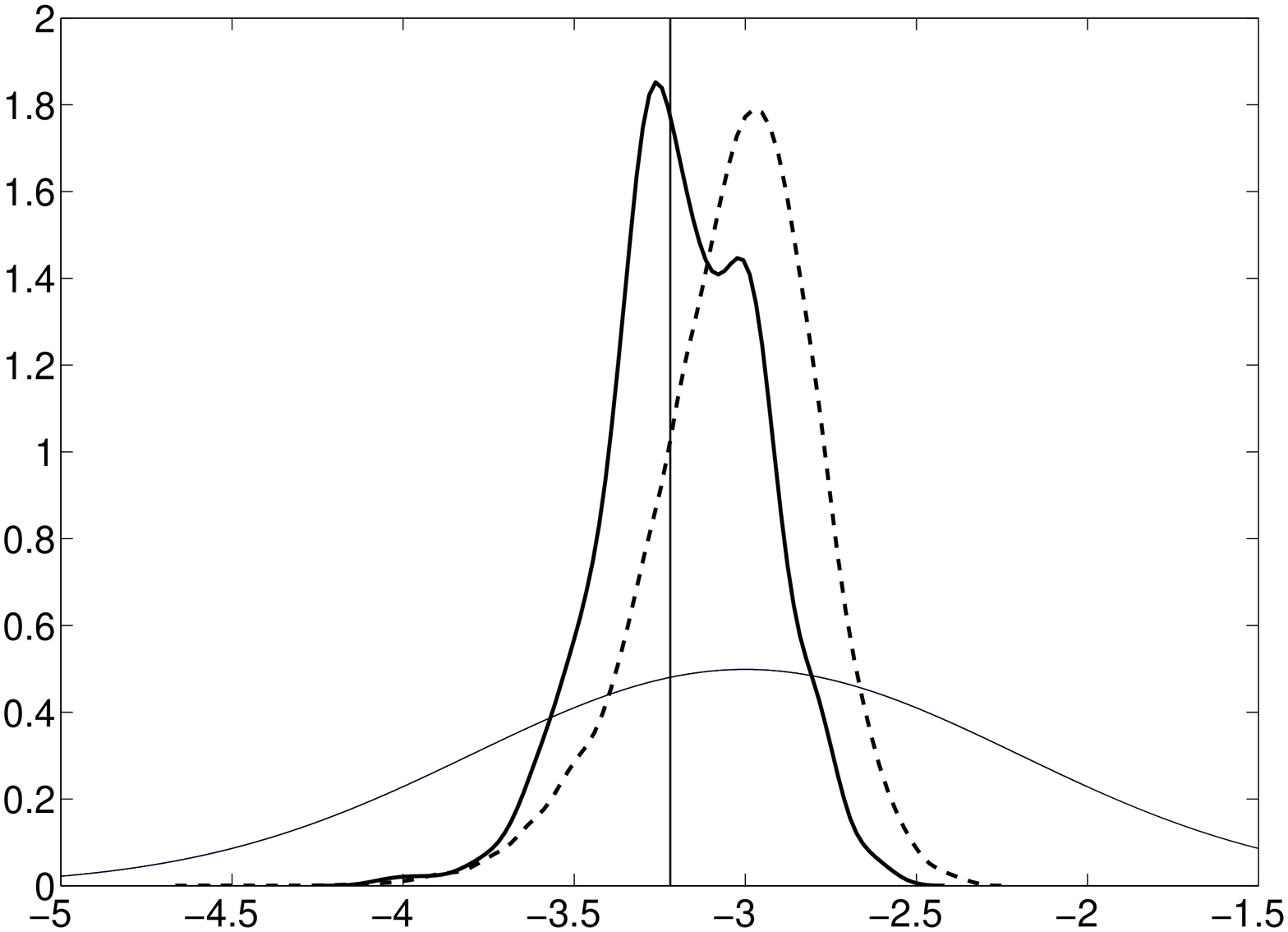}}
\caption{\footnotesize{Theophylline example: subset of marginal posterior densities from the ABC-MCMC output (bold solid lines) where summary statistics have been obtained via linear regression; marginal posteriors from PMCMC when $K=500$ (dashed lines) and prior densities $\pi(\theta_j)$ (wide Gaussian curves). True parameter values are marked with vertical lines.}}
\label{fig:kde-theophylline}
\end{figure}

We compare our results against PMCMC, that is we use sequential Monte Carlo (SMC) to approximate the data-likelihood and plug such approximation into a Metropolis-Hastings algorithm. The general methodology presented in \cite{andrieu2010particle} allow for posterior inference on $(\theta,\{X_t\})$, but here we are ultimately interested in posterior inference for $\theta$; in this sense we use a so-called ``marginal'' approach. As proved in \cite{andrieu2010particle}, despite using an approximation to the likelihood, the algorithm still target the exact posterior for any number of particles used. For particle generation we choose the ``bootstrap filter'' (see the classic monography by \cite{gordon2001sequential}) and used stratified resampling for particle randomization. We considered the same Gaussian priors $\pi(\theta_j)$ used for the ABC analysis. The PMCMC algorithm was run for 500,000 iterations with $K=100$ and 500 particles (we also experimented with $K=1,000$ however results didn't improve and are thus not reported). Results have been obtained in about 40 min with $K=100$ and 58 min with $K=500$ with an average acceptance rate of 30\% in both cases (we produced a carefully vectorised code for particles generation, i.e. for trajectories simulation, and since \textsc{Matlab} is a software well suited for vectorisation this explains the nonlinear time increase for increasing $K$). Same as for ABC we used the adaptive MCMC by \cite{haario-et-al(2001)} to propose parameters. Given the considered data-poor scenario (recall that only nine noisy measurements are available) it is not surprising to notice some difficulty in identifying the true parameters even with PMCMC, but this is due to the challenging experimental setup not to PMCMC. A subset of the posterior marginal densities obtained with PMCMC are compared in Figure \ref{fig:kde-theophylline} with those obtained with ABC when linear regression has been used. In order to ease comparison between methods the ESS have always been computed on samples having fixed lengths of 5,700 elements (which is the size of the chain considered for ABC-MCMC in this section).

\subsubsection*{Results obtained with Lasso}

We now consider results obtained when using lasso (\cite{tibshirani-1996}) for the estimation of ABC summary statistics during the ``training''. We used the same Gaussian priors $\pi(\theta_j)$ considered in the previous section and used lasso only during the training phase. Therefore we run the ABC-MCMC using summary statistics provided via lasso and used the same settings as in the previous analysis during the ABC-MCMC.
By inspecting plots similar to those in Figure \ref{fig:parameters-vs-bandw} (not reported) draws having bandwidth smaller than $\delta^*=0.09$ are retained, see Table \ref{tab:theophylline-estimates} for results. 
In this application lasso is outperformed by linear regression, which is a result in line with the finding in \cite{fearnhead-prangle(2011)}, see in particular the higher ESS values for linear regression and the values of posterior means when compared with PMCMC with $K=500$ (also notice that sometimes posterior intervals for lasso fail to include the true values of parameters). Nevertheless we use lasso in the application considered in section \ref{sec:stoch-kin-net} as in that case summary statistics have to be computed in a scenario often denoted in regression literature with ``$n\ll p$'', which can be handled with lasso.

In conclusion, having obtained sufficiently good results (at least with the linear regression approach) with the considered methodology is remarkable, given the several approximations involved, namely the use of numerical approximations to the SDE solution, the use of non sufficient statistics $\bm{S}(\cdot)$, a value for $\delta$ which is not zero and the limited amount of data $n$ subject to measurement error.

\subsection{Stochastic kinetic networks}\label{sec:stoch-kin-net}

Here we consider a more complex simulation study, involving a multidimensional SDE and using several experimental setups, namely fully and partially observed systems, with and without measurement error.
We study a model proposed in \cite{golightly-wilkinson-2010} for biochemical networks.
Consider the set of reactions $\{R_1,R_2,...,R_8\}$ defined by
\footnotesize
\begin{equation*}
\begin{aligned}
R_1 &: DNA+P_2\rightarrow DNA\cdot P_2 \qquad\qquad\\
R_3 &: DNA \rightarrow DNA+RNA \\ 
R_5 &: 2P \rightarrow P_2 \\ 
R_7 &: RNA \rightarrow \emptyset
\end{aligned}
\begin{aligned}
R_2 &: DNA\cdot P_2\rightarrow DNA + P_2\\
R_4 &: RNA \rightarrow RNA+P\\
R_6 &: P_2 \rightarrow 2P\\
R_8 &: P \rightarrow \emptyset.
\end{aligned}
\end{equation*}
\normalsize
These reactions represent a simplified model for prokaryotic auto-regulation based on the mechanism of dimers of a protein coded for by a gene repressing its own transcription. The ``reactants'' and ``products'' in a specific reaction in a model can be represented via the \textit{stoichiometry matrix}, including the reactions that a specific ``species'' ($RNA$, $P$, $P_2$, $DNA\cdot P_2$, $DNA$) is part of, and whether the species is a reactant or product in that reaction.

Consider the time evolution of the system as a Markov process with state $\bm{X}_t$, where $\bm{X}_t$ contains the number of molecules (i.e. non-negative integers) of each species at time $t$, i.e. $\bm{X}_t=(RNA,P,P_2,DNA\cdot P_2,DNA)^T$. In a stoichiometry matrix reactants appear as negative values and products appear as positive values. However in practice for many biochemical network models the presence of conservation laws lead to rank degeneracy of the stoichiometry matrix $\bm{S}$ and therefore redundant species should be eliminated from the model prior to conducting inference. In our case there is just one conservation law, $DNA\cdot P_2 + DNA = k$, where $k$ is the number of copies of this gene in the genome, which is reasonable to assume known.  We can simply remove $DNA \cdot P_2$ from the model, replacing any occurrences of $DNA\cdot P_2$ with $k-DNA$. Therefore we have the following stoichiometry matrix 
\[
\bm{S}=\left (
\begin{matrix}
0 & 0 & 1 & 0 & 0 & 0 & -1 & 0\\ 
0 & 0 & 0 & 1 & -2 & 2 & 0 & -1\\
-1 & 1 & 0 & 0 & 1 & -1 & 0 & 0 \\
-1 & 1 & 0 & 0 & 0 & 0 & 0 & 0\\
\end{matrix}
\right ).
\]
Now assume associated to each reaction $i$ a rate constant $c_i$ ($i=1,2,...,8$) and a ``propensity function'' $h_i(\bm{X}_t, c_i)$, such that $h_i(\bm{X}_t, c_i)dt$ is the probability of a type $i$ reaction occurring in the time interval $(t, t + dt]$.
A continuous time (diffusion) approximation to the discrete process $\bm{X}_t$ can be written as
\begin{equation}
d\bm{X}_t = \bm{S}\bm{h}(\bm{X}_t,\bm{c})dt+\bm{S}\sqrt{\mathrm{diag}(\bm{h}(\bm{X}_t,\bm{c}))}d\bm{W}_t
\label{eq:cle-alternative}
\end{equation}
where $\bm{X}_t=(RNA,P,P_2,DNA)^T$, $\bm{c} = (c_1,c_2,...,c_8)$ and we take $\bm{h}(\bm{X}_t, \bm{c}) = (c_1DNA\times P_2, c_2(k-DNA), c_3DNA, c_4RNA,
c_5P(P-1)/2, c_6P_2, c_7RNA,\allowbreak c_8P)^T$
(notice that here $\times$ denotes multiplication), $d\bm{W}_t =(dW_{t,1},\allowbreak ...,dW_{t,8})^T$ 
and the $dW_{t,i}$ are i.i.d. $dW_{t,i}\sim N(0,dt)$,  $i=1,2,...,8$. Object of the inference is the estimation of $\bm{c}$. 

Model \eqref{eq:cle-alternative} could have a more elegant formulation, see e.g. \cite{golightly-wilkinson-2010}, however the proposed version solves some numerical issues related to taking the square root of non-positive definite matrices (arising when trajectories simulated with a numerical approximation scheme turn negative). However even when taking such precautions the numerical discretization scheme still does not guarantee that our state vector will stay non-negative for all $t$, hence we take absolute values $\mathrm{diag}(|\bm{h}(\bm{X}_t,\bm{c})|)$ under the square root, as in \cite{higham(2008)}. Notice that \eqref{eq:cle-alternative} has the minor disadvantage of requiring a vector $d\bm{W}_t$ having dimension larger than $\bm{X}_t$, to preserve dimensionality, i.e. we need to consider an $8-$dimensional vector of Brownian increments whereas $\dim(\bm{X}_t)=5$, thus introducing unnecessary redundancy  in the problem formulation.

We used the same setup as in \cite{golightly-wilkinson-2010}, except where modifications were required, as discussed later on.
In all experiments data were produced using 
$(c_1,c_2,...,c_8)= (0.1,0.7,0.35,0.2,0.1, \allowbreak 0.9,0.3,0.1)$ and $k$ was assumed known and set to $k=10$.
The starting value of the system is $\bm{X}_0=(RNA_0,P_0,P_{2,0},DNA_0)=(8,8,8,5)$. Since the $c_j$'s represent rates the mathematical problem is reformulated in terms of $\log c_j$.
We consider several scenarios, namely fully observed systems with and without measurement error and partially observed systems with measurement error. For all simulation designs data are generated via the ``Gillespie algorithm'' \citep{gillespie(1977)} on the time interval $[0,49]$, to ensure exact simulation of the biochemical system defined via the reactions $\{R_1,...,R_8\}$ and the constant rates $c_j$ specified above. Then the simulated $\bm{X}_t$ (non-negative integers representing number of molecules) are
sampled at 50 integer times $\{0,1,2,...,49\}$ and such $\{\bm{X}_{t_i}\}_{i}$ represent our data, unless Gaussian measurement error is added, and in the latter case data are denoted with $\{\bm{y}_{t_i}\}_{i}$. However we conventionally denote data with $\{\bm{y}_{t_i}\}_{i}$ in all cases, to simplify the exposition.
The dimension of the measured $\bm{y}_i\in \mathbb{R}^{d_i}$ at a given sampling time $t_i$ is $d_i\leq 4$, with equality holding for all $i$ when considering a fully observed system. 
We use the Gillespie algorithm for data generation, but not when implementing our inferential method as it would result computationally costly; instead trajectories from the SDE are numerically generated via Euler-Maruyama using a constant stepsize $h=0.1$.

\subsubsection*{Fully observed system}

We consider two different setup for the fully observed scenario, where all the coordinates of $\bm{X}_t$ are observed at the same set of time-points: in the first setup data are generated without measurement error, i.e. $\varepsilon_{ij}\equiv 0$ ($i=0,1,...,n$, $j=1,...,d$) and we denote the obtained data with $\mathcal{D}_1$; (ii) in the second setup data are perturbed with measurement error $\varepsilon_{ij}\sim N(0,5)$ independently for every $i$ and for each coordinate $j$ of the state vector $\bm{X}_t$. In (ii) we assume the variance $\sigma^2_{\varepsilon}=5$ to be known and denote obtained data with $\mathcal{D}_2$.
We consider $\mathcal{D}_1$ first:  
the vector parameter object of our inference is $\bm{\theta}=(\log c_1,\log c_2,...,\log c_8)$.

\begin{figure}[h]
\centering
\subfloat[Subfigure 1 list of figures text][$\log c_4$ vs bandwidth]{
\includegraphics[width=0.4\textwidth]{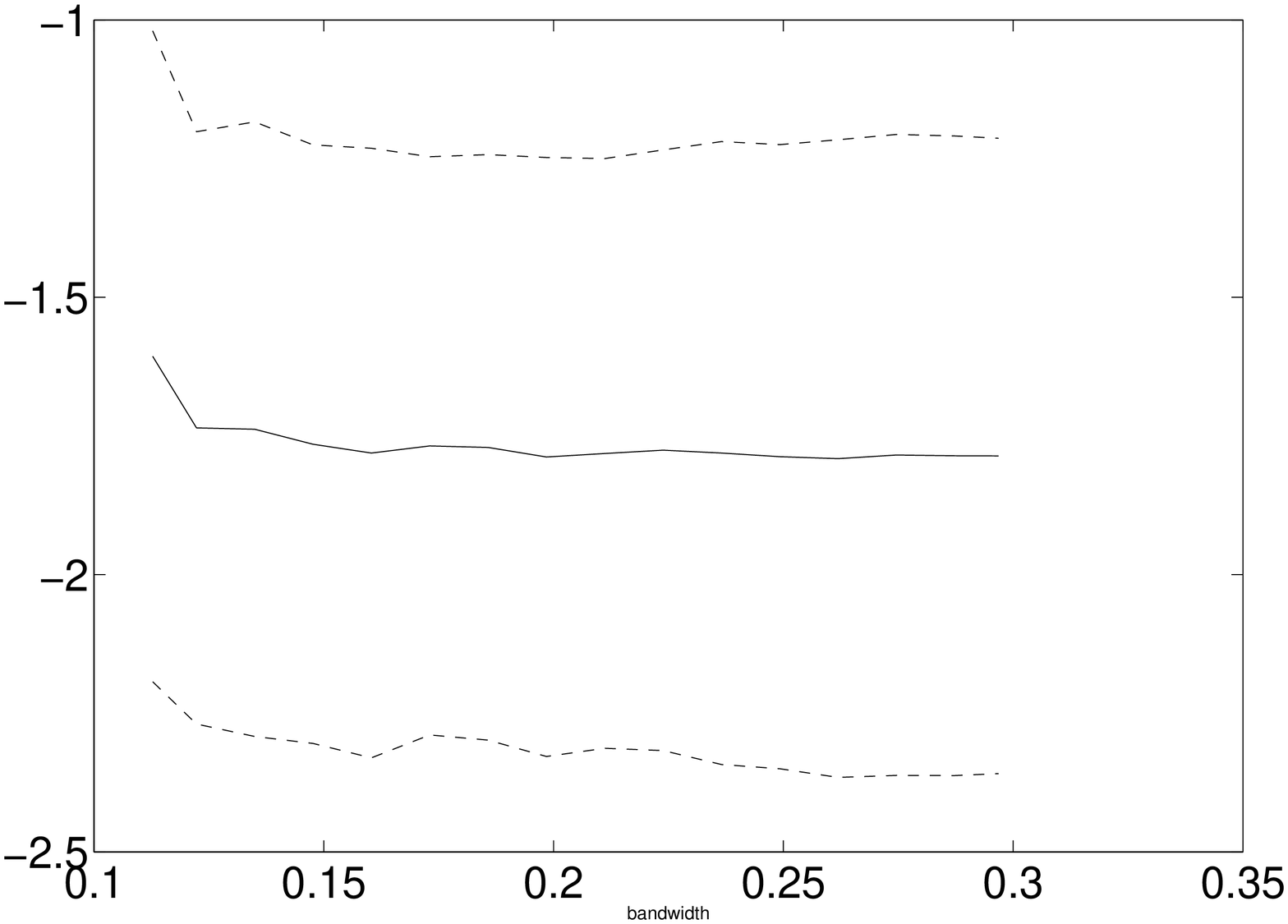}}
\qquad
\subfloat[Subfigure 2 list of figures text][$\log c_8$ vs bandwidth]{
\includegraphics[width=0.4\textwidth]{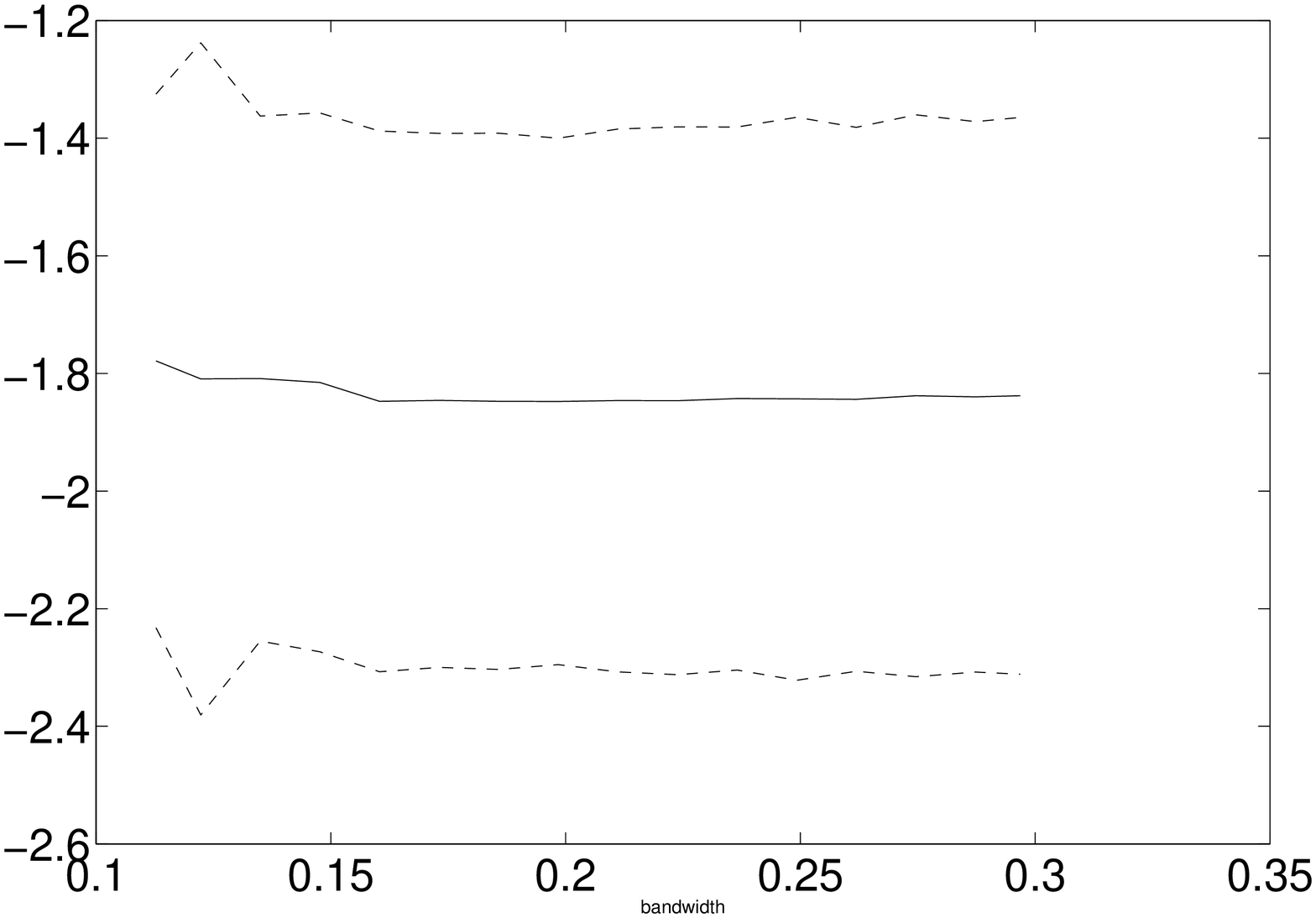}}\\
\caption{\footnotesize{Stochastic networks example using $\mathcal{D}_1$: posterior means from ABC-MCMC [$\pm 2$ SD] vs bandwidth (bandwiths are on the abscissas).}}
\label{fig:stockntw-parameters-vs-bandw}
\end{figure}

We defined the following priors $\pi(\bm{\theta})$: $\log c_1 \sim N(-2.6,0.25^2)$, $\log c_2 \sim N(-0.6,0.4^2)$, $\log c_3 \sim N(-1.5,0.4^2)$, $\log c_4 \sim N(-1.8,0.4^2)$, $\log c_5 \sim N(-2.4,0.2^2)$, $\log c_6 \sim N(-1,0.4^2)$, $\log c_7 \sim N(-1.85,0.3^2)$, $\log c_8 \sim N(-1.8,0.3^2)$. 
Prior to starting the ABC-MCMC we need to compute the statistics $\bm{S}(\cdot)$ with lasso, and this is accomplished using \texttt{simpar=25x200=5,000} simulations. 
Algorithm \ref{alg:lfmcmc-earlyrej} is executed for 3 million iterations, with a thinning of 50 values. We set for the bandwidth a truncated exponential prior $\delta\sim \mathrm{Exp}(0.07)$, a starting value equal to 0.2 and $\delta_{\mathrm{max}}=0.3$, resulting in an acceptance rate of about 6.5\% using adaptive Metropolis random walk. 
A burn-in of 150,000 draws (corresponding to 3,000 draws discarded after thinning) is considered. For $\mathcal{D}_1$ an analysis of the bandwidth effect on the posterior means is given in Figure \ref{fig:stockntw-parameters-vs-bandw}, and for illustration purposes only plots pertaining to $\log c_4$ and $\log c_8$ are reported.

\begin{table}[h!]
\centering \scriptsize
\begin{tabular}{ccccc}
\hline
\hline
{} & True parameters & $\mathcal{D}_1$ & $\mathcal{D}_2$ & $\mathcal{D}_3$ \\ 
\hline
$DNA_0$  & 5      &      ---              &    ---             &      5.90  \\
         &        &                       &                    &      [4.550, 7.627]  \\ 
$c_1$    & 0.1    &      0.074            &     0.072          &       0.074\\
{}       &        &      [0.052, 0.108]   &    [0.051, 0.104]  &        [0.049, 0.111] \\
$c_2$    & 0.7    &       0.526           &     0.583          &        0.524\\
{}       &        &      [0.352, 0.773]  &    [0.378, 0.944]  &       [0.288, 0.941]\\
$c_1/c_2$& 0.143  &    0.142        &            0.124         &       0.141 \\ 
{}       &        &                 &                & \\
$c_3$    & 0.35 &      0.208       &             0.201         &        0.205\\ 
{}       &        &    [0.130, 0.316]    &     [0.124, 0.313]       & [0.127, 0.343]\\
$c_4$    & 0.2 &       0.172              &        0.178       &        0.175 \\ 
{}       &        &    [0.102, 0.296]    &      [0.106, 0.295] &        [0.102, 0.276]\\
$c_5$    & 0.1 &       0.087       &              0.086       &         0.088 \\ 
{}       &        &    [0.064, 0.121]  &        [0.061, 0.119]          & [0.066, 0.116] \\
$c_6$    & 0.9 &       0.419      &               0.421               & 0.393\\ 
{}       &        &    [0.295, 0.611]     &      [0.282, 0.627]          & [0.265, 0.589]\\
$c_5/c_6$&     0.111   &     0.208        &       0.204            &    0.224\\
{}       &             &      &                                    &     \\
$c_7$    & 0.3 &       0.165      &               0.180            &    0.169\\ 
{}            &        & [0.113, 0.254]     &    [0.118, 0.277]    &    [0.113, 0.258]\\
$c_8$    & 0.1 &       0.161      &               0.158            &   0.154\\ 
{}            &        & [0.100, 0.262]    &      [0.104, 0.241]   &    [0.089, 0.228]\\
$\sigma_{\varepsilon}$ & 2.236  &  --- &                ---        &   3.800 \\ 
{}            {} &         & {}            &                   {}         & [2.566, 5.779] \\
\hline
\end{tabular}
\caption{\footnotesize{Stochastic networks example: posterior means from the ABC-MCMC output (first line) and 95\% central posterior intervals (second line). Depending on the simulation assumptions a (---) means that for the given data set the parameter is not part of the model or has been treated as a known constant. See main text for details.}}
\label{tab:stochnetw-estimates} 
\end{table}

Variation in the posterior means are evident within the range allowed for $\delta$ during the simulation, i.e. in the interval [0,0.3]. Although theoretically we could  conduct inference based on a very small value for $\delta$, in practice it is not safe to consider only those $\bm{\theta}_r$ corresponding to very small $\delta$'s as in practice these imply a poor mixing. It is important to make use of plots of the type given in Figures \ref{fig:parameters-vs-bandw} and \ref{fig:stockntw-parameters-vs-bandw} to detect sharp increases/decreases in the posterior means. We want results corresponding to ``small'' $\delta$'s but \textit{not only} those, as they result often associated with low--mixing regions and, as such, regions where the exploration of the posterior distribution is poor.
We therefore filter the chains and keep the draws corresponding to relatively small $\delta$'s (for statistical precision) while allowing for larger $\delta$'s as long as the posterior means look stable for varying $\delta$: we ultimately select draws having $\delta< 0.17$. This provides us with about 4,750 draws which we use for posterior inference. Results for the relevant parameters are given in Table \ref{tab:stochnetw-estimates}. Notice that, as in \cite{golightly-wilkinson-2010}, the estimates for $c_2$ is far from being excellent, however we are able to recover the ratio $c_1/c_2$ which represents ``propensity for the reversible repression reactions''. The ESS for the eight parameters range from 206 to 379 (median ESS is 310), that is from 4.3\% to 7.9\% of the considered 4,750 draws.
Using the same computer as in the previous example it requires 6.7 hrs to run algorithm \ref{alg:lfmcmc-earlyrej}, that is generating the 3,000,000 draws with a \textsc{Matlab} implementation.
 
For data $\mathcal{D}_2$, where observations are affected with Gaussian zero-mean measurement error with known variance $\sigma_{\varepsilon}^2=5$, we followed a similar procedure. 
No striking differences emerge with respect to the results obtained on $\mathcal{D}_1$, which is a result in itself, as the method seems able to filter out the measurement error variability by returning estimates similar to those obtained with measurement-error-free data.

\subsubsection*{A partially observed system}

Here we consider a further data set denoted with $\mathcal{D}_3$ having the $DNA$ coordinate unobserved. That is we still consider $\bm{X}_t=(RNA_t,P_t,P_ {2,t},DNA_t)^T$ and only the first three coordinates are measured (at the same 50 integer times as previously described) and therefore at any sampling time $t=t_i$ we have measurements for $(RNA_{i},P_i,P_{2,i})^T$, that is $\bm{y}_i\in \mathbb{R}^3_{+}$ and a total of $150$ observations is available. We assume that at any $t_i$ each coordinate of $(RNA,P,P_2)_i^T$ is perturbed with independent homoscedastic noise, $\bm{y}_i=(RNA_i,P_i,P_{2,i})^T+\bm{\varepsilon}_i$ with $\bm{\varepsilon}_i\sim N_3(\bm{0},\sigma^2_{\varepsilon}\bm{I}_3)$, however this time $\sigma_{\varepsilon}$ is considered unknown and has to be estimated. Since $DNA$ is not measured, we set a Gaussian prior on $\log DNA_0\sim N(1.8,0.16^2)$ and for the error variability we set $\log \sigma_{\varepsilon}\sim N(1.4,0.25^2)$. We considered $\delta\sim \mathrm{Exp}(0.05)$ and $\delta_{\mathrm{max}}=0.4$.
Because of the uncertainty placed on $DNA_0$ we generated a longer chain of 5,000,000 draws in 10.5 hrs of computation and obtained an average acceptance rate of 6.8\%. We retained about 7,450 draws corresponding to $\delta<0.15$, see  Table \ref{tab:stochnetw-estimates} for results. Interestingly, posterior means do not result considerably different from the previous attempts with $\mathcal{D}_1$ and $\mathcal{D}_2$, despite the increased uncertainty in the simulation setup. The initial state $DNA_0$ is correctly identified but residual variation $\sigma_\varepsilon$ is not. We computed the ESS on a sample of size 4,750, same as for $\mathcal{D}_1$: ESS values range from 65 to 348 (median ESS is 202).

\section{Summary}

We considered approximate Bayesian computation (ABC) to perform inference for complex, multidimensional, partially observed stochastic dynamical systems subject to measurement error. 
A simple risk--free modification to a standard ABC-MCMC algorithm is proposed, reducing the computational effort by 40--50\% in our experiments and thus easing the practical application of ABC methodology for stochastic modelling. The setup can be largely automated, thanks to recent results on the determination of informative summary statistics as proposed in \cite{fearnhead-prangle(2011)}.
As shown in our simulation studies, results are encouraging, and this is comforting since ``exact'' methods (or Monte Carlo approximations thereof) are rarely feasible under complex multidimensional scenarios where the computational requirements are still highly demanding. The resulting methodology is flexible and is not limited to the state-space modelling framework. The provided \texttt{abc-sde} \textsc{Matlab} package can be considered to fit stochastic models with latent dynamics expressed via multidimensional SDE models using ABC-MCMC, see \url{http://sourceforge.net/projects/abc-sde/}. Alternatives based on sequential Monte Carlo (SMC) methods are available, notably \cite{toni-et-al-2009} and \cite{liepe2010abc} which have shown excellent inferential performance as well as the possibility to avoid completely the introduction of summary statistics. For applications larger than the ones here discussed and when using commonly available computational devices without using e.g. GPU computing, we had practical difficulties when trying to apply our own implementations of SMC-based methods (both for exact and ABC inference) to fit several hundreds or thousands of observations. However SMC methods are well suited for parallelization, see \cite{liepe2010abc} and \cite{libbi}.

This work should be particularly useful to modellers dealing with multidimensional stochastic systems, rather than experimenting with one-dimensional models without measurement error for which a number of theoretical and computational results are available (see the review papers by \cite{sorensen(2004)} and \cite{hurn-jeisman-lindsay(2007)}). On the other side we attempted at performing inference for multidimensional SDE models incorporating measurement error, which is an unavoidable (and non-trivial) task to seriously study increasingly large models for real world applications.

\bigskip

\textbf{Acknowledgments:} this work has been partially supported by the Faculty of Science at Lund University under the grant ``Money Tools'' (verktygspengar). The author wishes to acknowledge the comments provided by an associate editor and two reviewers which improved the quality of this work considerably.

\bibliography{biblio}
\bibliographystyle{plainnat}

\end{document}